\documentclass[sigconf]{acmart}

\usepackage{booktabs} 
\usepackage{url}
\copyrightyear{2017} 
\acmYear{2017} 
\setcopyright{rightsretained}
\acmConference{SIGIR'17}{}{August 7--11, 2017, Shinjuku, Tokyo, Japan.}
\acmDOI{10.1145/3077136.3084370}
\acmISBN{ACM ISBN 978-1-4503-5022-8/17/08}

\begin{document}

\title{Joint Workshop on Bibliometric-enhanced Information Retrieval and 
Natural Language Processing\\ for Digital Libraries (BIRNDL 2017)}




\author{Muthu Kumar Chandrasekaran}
\affiliation{%
 \small
 \institution{School of Computing,\\National University of Singapore, Singapore} 
}
\email{muthu.chandra@comp.nus.edu.sg}

\author{Kokil Jaidka}
\affiliation{%
\small
 \institution{School of Arts \& Sciences,\\University of Pennsylvania, USA}
}
\email{jaidka@sas.upenn.edu}

\author{Philipp Mayr}
\affiliation{%
\small
 \institution{GESIS -- Leibniz Institute for the Social Sciences, Germany}
}
\email{philipp.mayr@gesis.org}

\begin{abstract}
The large scale of scholarly publications poses a challenge for scholars in 
information seeking and sensemaking. Bibliometrics, information retrieval~
(IR), text mining and NLP techniques could help in these search and look-up 
activities, but are not yet widely used. 
This workshop is intended to stimulate IR researchers and digital library 
professionals to elaborate on new approaches in natural language processing, 
information retrieval, scientometrics, text mining and recommendation 
techniques that can advance the state-of-the-art in scholarly document 
understanding, analysis, and retrieval at scale. The BIRNDL workshop at SIGIR 
2017 will incorporate an invited talk, paper sessions and the third edition 
of the Computational Linguistics (CL) Scientific Summarization Shared Task.

\end{abstract}



\maketitle

\section{Introduction} 
Over the past several years, the BIRNDL workshop and its parent workshops are 
establishing themselves as the primary interdisciplinary venue for the cross-
pollination of bibliometrics and information retrieval (IR) \cite{Mayr2015}. 
Our motivation as organizers of the workshop started from the observation 
that both communities share only a partial overlap; yet, the main discourse 
in both fields consists of different approaches to solve similar problems. We 
believe that a knowledge transfer would be profitable for both sides. A good 
overview of the symbiotic relationship that exists among bibliometrics, IR 
and natural language processing (NLP) has been presented by Wolfram 
\cite{Wolfram16}. A report of the past BIRNDL workshop has been published 
recently in The SIGIR Forum \cite{Cabanac2016}.

The goal of the BIRNDL workshop at SIGIR is to engage the IR community about 
the open problems in academic search. 
Academic search refers to the large, cross-domain digital repositories which 
index research papers, such as the ACL Anthology, ArXiv, ACM Digital Library, 
IEEE database, Web of Science and Google Scholar. 
Currently, digital 
libraries collect and allow access to papers and their metadata ---
including citations --- but mostly do not analyze the items they index. The 
scale of scholarly publications poses a challenge for scholars in their 
search for relevant literature. Finding relevant scholarly literature is the key 
theme of BIRNDL and sets the agenda for tools and approaches to be 
discussed and evaluated at the workshop.

Papers at the $2^{nd}$ BIRNDL workshop will incorporate insights from IR, 
bibliometrics and NLP to develop new techniques to address the open problems 
such as evidence-based searching, measurement of research quality, relevance 
and impact, the emergence and decline of research problems, identification of 
scholarly relationships and influences and applied problems such as language 
translation, question-answering and summarization. We will also address the 
need for established, standardized baselines, evaluation metrics and test 
collections. Towards the purpose of evaluating tools and technologies 
developed for digital libraries, we are organizing the $3^{rd}$ CL-SciSumm 
Shared Task based on the CL-SciSumm corpus, which comprises over 500 
computational linguistics (CL) research papers, interlinked through 
a citation network.

The organizers of the $2^{nd}$ BIRNDL workshop at SIGIR 2017\footnote{\url{http://wing.comp.nus.edu.sg/birndl-sigir2017/}} have previously 
organized other workshop series at premier IR and CS venues 
- notably, the 
Bibliometric-enhanced Information Retrieval (BIR) workshops in 2014, 2015 and 
2016 at ECIR~\cite{Mayr2016} and the NLPIR4DL workshop at ACL-IJCNLP (2009). 
Most recently, the BIRNDL workshop and the $2^{nd}$ CL-SciSumm Shared Task 
were co-located with JCDL 
2016\footnote{\url{http://wing.comp.nus.edu.sg/birndl-jcdl2016/}} 
\cite{Cabanac2016}, where 10 research papers and 10 system papers were 
presented\footnote{\url{http://ceur-ws.org/Vol-1610/}} (acceptance rate: 
30\%). 
In 2017, the BIRNDL workshop takes this legacy forward 
with a focus on scholarly publications and data, 
and an updated scientific summarization 
Shared Task for its participants.

This workshop will be relevant to scholars in computer and information 
science, specializing in IR and NLP. It will also be of importance to all 
stakeholders in the publication pipeline: 
practitioners,
publishers and policymakers. 
Today's publishers continue to provide new ways to support 
their consumers in disseminating and retrieving the right published works to 
their audience. 
Formal citation metrics are increasingly a factor in 
decision-making by universities and funding bodies worldwide, making the need 
for research in applying these metrics more pressing.

\section{Workshop Topics and Format}
Our goal is to encourage insights from IR, NLP and CL 
for scholarly document understanding, document analysis and retrieval in 
digital libraries. 
The papers presented at the workshop will touch upon 
several topics, including (but not limited to) full-text analysis, multimedia 
and multilingual analysis and alignment as well as the application of 
citation-based NLP, information retrieval and information seeking techniques 
in digital libraries. More specifically, our fields of interests include:
\begin{itemize}
\item Infrastructures for scientific text mining and IR
\item Semantic and Network-based indexing, navigation, searching and browsing 
in structured data
\item Discourse structure identification and argument mining from scientific 
papers
\item Summarization and question-answering for scholarly DLs
\item Recommendation for scholarly papers, reviewers, citations and 
publication venues
\item Measurement and evaluation of quality and impact
\item Metadata and controlled vocabularies for resource description and 
discovery; automatic metadata discovery, such as language identification
\item Disambiguation issues in scholarly DLs using NLP or IR techniques; data 
cleaning and data quality.
\end{itemize}

\subsection{Tentative Schedule of Events}
The workshop will start with a keynote titled \textit{``Do ``Future Work" 
sections have a real purpose? Citation links and entailment for global 
scientometric questions''} by Dr. Simone Teufel (University of 
Cambridge).  This session will be followed by regular research paper 
presentations, overview papers and posters on the Shared Task. 

\subsection{The CL-SciSumm Shared Task}
\label{sec:shared}

The $3^{rd}$ Computational Linguistics (CL) Scientific Summarization Shared 
Task, sponsored by Microsoft Research Asia, will be conducted as a part 
of this workshop. This is the first medium-scale shared task on scientific 
document summarization in the CL domain. It follows up 
on and extends the successful CL Shared Tasks conducted as a part of  
BIRNDL 2016~\cite{Cabanac2016}, and within the BiomedSumm Track 
at the Text Analysis Conference 2014 
(TAC~2014)~\cite{jaidka2014computational}. 
In the CL-SciSumm 2016~\cite{jaidka2016overview} Shared Task, fifteen teams 
from six countries signed up, and ten teams ultimately submitted and 
presented their results. 

The Shared Task comprises three sub-tasks in automatic research paper 
summarization on a new corpus of research papers, as described below.

Given: A topic consisting of a Reference Paper (RP) and up to ten Citing 
Papers (CPs) that all contain citations to the RP. Citations in the CP are 
pre-identified as the text spans (i.e., citances), that cite the RP.

\noindent\textbf{Task 1a:} For each citance, identify the spans of text 
(cited text spans) in the RP that most accurately reflect the citance.\\
\textbf{Task 1b:} For each cited text span, identify what facet of the 
paper it belongs to, from a predefined set of facets.\\
\textbf{Task 2} (optional bonus task): Finally, generate a structured 
summary of the RP from the cited text spans of the RP. The length of 
the summary should not exceed 250 words.\\
\textbf{Evaluation:} Task 1 will be scored by overlap of text spans 
measured by number of sentences in the system output vs gold standard. 
Task 2 will be scored using the ROUGE family of metrics between the 
system output, and i) human summaries, ii) community summaries comprising 
the cited text spans, and ii) the Abstract section of the reference paper.

This task is continues to be of interest to a broad community including 
those working in CL and NLP, especially in the sub-disciplines of text 
summarization, discourse structure in scholarly discourse, paraphrase, 
textual entailment and text simplification.

\section{Outlook}
This workshop is the first step to foster a reflection on 
interdisciplinarity, and the benefits that the disciplines Bibliometrics, IR 
and NLP can derive from it in the Digital Libraries context. 
The authors of 
accepted papers will be invited to submit extended versions of their work to 
the International Journal on Digital Libraries (IJDL). As an output of BIRNDL 
2016, a special issue of IJDL on ``Bibliometrics, Information Retrieval and 
Natural Language Processing in Digital Libraries'' is currently in 
preparation. In the future, we plan to continue to host this series of 
workshops and Shared Tasks at prominent IR, NLP and Digital Library venues.

\vspace{-.15cm}
\section*{Acknowledgments}
\small
We thank Microsoft Research Asia for their generous support in funding the 
development, dissemination and organization of the CL-SciSumm dataset and 
the Shared Task.
We are also grateful to the co-organizers of the $1^{st}$ BIRNDL 
workshop - Guillaume Cabanac, Ingo Frommholz, Min-Yen Kan and Dietmar 
Wolfram, for their continued support and involvement.

\vspace{-.25cm}
\bibliographystyle{ACM-Reference-Format}
\bibliography{sigproc}


\end{document}